\documentclass[12pt,preprint]{aastex}
\usepackage{graphicx}
\usepackage{epstopdf}
\usepackage{apjfonts}
\usepackage[usenames,dvips]{color}
\usepackage{longtable}
\usepackage{amsmath}
\usepackage{threeparttable}

\usepackage[bookmarks,breaklinks]{hyperref}
   \hypersetup{
     colorlinks,
     citecolor=blue,
     linkcolor=blue,
    }
\begin{document}
%
%
\shorttitle{1.3 mm VLBI OBSERVATIONS OF 1924-292}
\shortauthors{Lu et al.}
\title{RESOLVING THE INNER JET STRUCTURE OF 1924-292 WITH THE EVENT HORIZON TELESCOPE}
\author{Ru-Sen Lu \altaffilmark{1},
          Vincent L.\ Fish\altaffilmark{1},
          Jonathan Weintroub\altaffilmark{2},
          Sheperd S.\ Doeleman\altaffilmark{1}, 
          Geoffrey C.\ Bower\altaffilmark{3},
         Robert Freund\altaffilmark{4},
         Per Friberg\altaffilmark{5},
          Paul T. P. Ho\altaffilmark{6},
         Mareki Honma\altaffilmark{7,8},
         Makoto Inoue\altaffilmark{6},
          Thomas P.\ Krichbaum\altaffilmark{9},
          Daniel P.\ Marrone\altaffilmark{4},
         James M.\ Moran\altaffilmark{2},
         Tomoaki Oyama\altaffilmark{7},
          Richard Plambeck\altaffilmark{3},
         Rurik Primiani\altaffilmark{2},
         Zhi-Qiang Shen\altaffilmark{10,11},
         Remo P.\ J.\ Tilanus\altaffilmark{5,12},
         Melvyn Wright\altaffilmark{3},
          Ken H.\ Young\altaffilmark{2},
          Lucy M.\ Ziurys\altaffilmark{4},
          J. Anton Zensus\altaffilmark{9}
}
\email{rslu@haystack.mit.edu}
\altaffiltext{1}{Massachusetts Institute of Technology, Haystack
  Observatory, Route 40, Westford, MA 01886, USA}
 \altaffiltext{2}{Harvard-Smithsonian Center for Astrophysics, 60 Garden
   St., Cambridge, MA 02138, USA}
 \altaffiltext{3}{University of California Berkeley, Dept.\ of
   Astronomy, Radio Astronomy Laboratory, 601 Campbell Hall, Berkeley, CA
   94720-3411, USA}
  \altaffiltext{4}{Arizona Radio Observatory, Steward Observatory,
   University of Arizona, 933 North Cherry Ave., Tucson, AZ 85721-0065,
   USA}
 \altaffiltext{5}{James Clerk Maxwell Telescope, Joint Astronomy Centre,
   660 North A'ohoku Place, University Park, Hilo, HI 96720, USA}
   \altaffiltext{6}{Institute of Astronomy and Astrophysics, Academia
   Sinica, P.O. Box 23-141, Taipei 10617, Taiwan}
 \altaffiltext{7}{National Astronomical Observatory of Japan, Osawa
   2-21-1, Mitaka, Tokyo 181-8588, Japan}
\altaffiltext{8}{The Graduate University for Advanced Studies, Osawa 2-21-1, Mitaka, Tokyo, 181-8588, Japan}
 \altaffiltext{9}{Max-Planck-Institut f\"{u}r Radioastronomie, Auf dem
   H\"{u}gel 69, D-53121 Bonn, Germany }
  \altaffiltext{10}{Key Laboratory for Research in Galaxies and Cosmology, Shanghai Astronomical Observatory, Chinese Academy of Sciences, 80
Nandan Rd, Shanghai 200030, China}
\altaffiltext{11}{Key Laboratory of Radio Astronomy, Chinese Academy of Sciences, China}
 \altaffiltext{12}{Netherlands Organization for Scientific Research, Laan van Nieuw Oost-Indie 300, NL2509 AC The Hague, The Netherlands}

\begin{abstract}
We present the first 1.3\,mm (230\,GHz) very long baseline interferometry model image of an AGN jet using closure phase techniques with a four-element array. The model image of the quasar 1924-292 was obtained with four telescopes at three observatories: the James Clerk Maxwell Telescope (JCMT) on Mauna Kea in Hawaii, the Arizona Radio Observatory's Submillimeter Telescope (SMT) in Arizona, and two telescopes of the Combined Array for Research in Millimeterwave Astronomy (CARMA) in California in April 2009. With the greatly improved resolution compared with previous observations and robust closure phase measurement, the inner jet structure of 1924-292 was spatially resolved. The inner jet extends to the northwest  along a position angle of $-53^\circ$ at a distance of 0.38\,mas from the tentatively identified core, in agreement with the inner jet structure inferred from lower frequencies, and making a position angle difference of $\sim 80^{\circ}$ with respect to the cm-jet. The size of the compact core is 0.15\,pc with a brightness temperature of $1.2\times10^{11}$\,K.  Compared with those measured at lower frequencies, the low brightness temperature may argue in favor of the decelerating jet model or particle-cascade models. The successful measurement of closure phase paves the way for imaging and time resolving Sgr A* and nearby AGN with the Event Horizon Telescope.

\end{abstract}
\keywords{galaxies: active - galaxies: jets - quasars: individual (1924-292) - radio continuum: general - techniques: high angular resolution - techniques: interferometric}

\section{Introduction}
The quasar 1924-292 (PKS 1921-293, OV-236) is one of the brightest and most compact flat-spectrum radio sources in the sky. It has been classified as an optically violent variable \citep{1981Natur.289..384W,1988AJ.....96.1215P} and highly polarized quasar \citep{1990ApJ...360..396W}. As a radio-loud blazar, it shows strong variability from radio to X-ray. This source is also included in the Fermi-LAT 1-year Point Source Catalog \citep{2010ApJS..188..405A}. At its redshift of z=0.352 \citep{1981Natur.289..384W}, an angular resolution of 1\,mas corresponds to 4.93 pc ($H_{\rm 0}= 71 \rm km s^{-1}  Mpc^{-1}$, $\Omega_{\rm M} = 0.27$ and $\Omega_\Lambda = 0.73$).

1924-292 is completely unresolved with Very Large Array (VLA) observations made at 6\,cm and 20\,cm \citep{1985AJ.....90..846D,1982AJ.....87..859P}. Very long baseline interferometry (VLBI) observations made at cm wavelengths show a typical core-jet structure with the jet extending about 10 mas to northeast along a position angle (P.A.) of approximately $25^{\circ}$--$30^{\circ}$\citep[e.g.,][]{1998AJ....115.1295K,1989AJ.....98....1P,1997AJ....114.1999S,1998ApJ...497..594T}. With increased resolution, VLBI observations at 7 and 3.5~mm and VLBI space observatory program observations showed that the inner jet curves sharply and is oriented toward the northwest (extends up to about 1 mas) with a time-varying position angle \citep{2008AJ....136..159L,1999PASJ...51..513S,2002aprm.conf..401S}. \citet{2002aprm.conf..401S} reported that superluminal motion (about 3\,c) has been detected. The innermost ($<$ 1\,pc) region, however, was characterized by two equally compact components whose relative positions were unchanged over about 6.5 years covered by their observations. At cm wavelengths, this source also has one of the highest brightness temperatures of $\gtrsim 3 \times 10^{12}$ K measured in the source rest frame, in excess of the inverse Compton limit for synchrotron radiation \citep{1989ApJ...336.1105L,1996AJ....111.2174M,1999PASJ...51..513S,1998ApJ...497..594T}.

VLBI observations at short millimeter wavelengths ($\lambda \leq$1.3\,mm, $\nu \geq 230$\,GHz) have traditionally been challenging due to the limited sensitivity of  the instruments and  atmospheric phase fluctuations. With the application of new technical developments (such as phased-array processors, wide-bandwidth digital backends, and high-data-rate recorders) and the appearance of new suitable antennas, recent observations have established the technical feasibility of VLBI at short millimeter wavelengths and have opened a new window to directly study and image black holes with the Event Horizon Telescope (EHT)~\citep{Doeleman2008,2011ApJ...727L..36F}. 

Closure phase, which is the sum of the interferometric phase around a triplet of antennas, is largely immune to atmospheric and instrumental complex gain variations~\citep{1974ApJ...193..293R,1984ARA&A..22...97P}. It is closely related to the asymmetry of the emission and thus is a robust observable for understanding source structures with high resolution. In the low signal-to-noise (SNR) regime, however, it has been inherently difficult to obtain closure phase with high-frequency VLBI at 1.3\,mm. With the deployment of new VLBI systems and enhancement of software capabilities, closure phases have now been robustly measured by the EHT on the quasar 1924-292, allowing us to model the compact structure on submilliarcsecond scales.

\section{OBSERVATIONS AND DATA REDUCTION}
\label{observations}
\begin{figure}[h!]
\begin{center}
\raisebox{0.75cm}{\includegraphics[width=0.475\textwidth,clip]{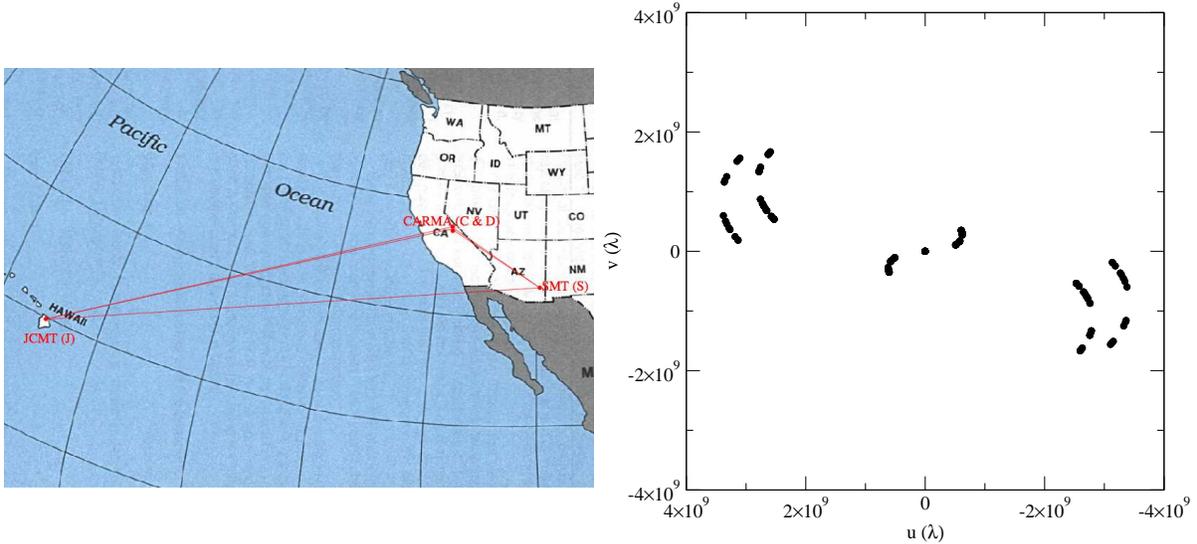}}
\includegraphics[width=0.475\textwidth,clip]{fg1b.eps}
\caption{Left: Map showing the four elements of the array for the 1.3\,mm VLBI observations on 1924-292. Right: Plot of the uv-coverage.}
\label{Fig:fg1}
\end{center}
\end{figure}

 On 5--7 April 2009 (days 95--97), 1924-292 was observed as a calibration source during observations of the Galactic center at 1.3\,mm with a four-station VLBI array consisting of the James Clerk Maxwell Telescope (JCMT; hereafter J) on Mauna Kea in Hawaii, the Arizona Radio Observatory's Submillimeter Telescope (SMT; hereafter S) in Arizona, and two telescopes of the Combined Array for Research in Millimeterwave Astronomy (CARMA; hereafter C\&D, located 60 m apart) in California \citep{2011ApJ...727L..36F}.  The JCMT made use of the eSMA infrastructure, with the Submillimeter Array providing the hydrogen maser standard and first local oscillator signal and housing the digital backend and data recorders. Figure~\ref{Fig:fg1} (left) shows the four elements of the array. The observations of 1924-294 comprise five-10\,minute scans each day with uv coverage shown in Figure~\ref{Fig:fg1} (right), resulting in a synthesized beam size of $\sim 0.26 \times 0.06$ mas with a position angle of $-20^{\circ}$. Observations were performed in LCP in two 480-MHz bands centered at 229.089 and 229.601 GHz (low and high bands) with an aggregate data rate of 3.84 Gigabit/sec (2-bit sampling) at each site. The data correlation was performed at MIT Haystack Observatory in Westford, Massachusetts, on the Mark4 VLBI correlator.
 
At short millimeter wavelengths, fluctuations in the atmospheric path length cause a loss of coherence in the VLBI data.
After data correlation, atmospheric coherence times were determined from the variation in coherence with integration time.
A search for a peak in SNR in delay and delay rate space was performed for each scan, after segmenting the data at the coherence time and incoherent averaging over the full scan length.
The segmented data were then incoherently averaged at the optimal values of delay and rate from the fringe search to reconstruct the visibility amplitude, during which corrections were introduced to account for the noise bias. The segmentation time is short enough that the amplitude information was preserved. We adopted a coherence time ranging from 3 to 5\,s (depending on the observing night). The coherence time is shorter than the coherent integration time corresponding to a 5\,\% amplitude loss for all scans, and coherence losses are limited to negligible levels. The segmented bispectrum (triple products of the complex visibilities) was also formed, and averaged closure phases were determined as an average of the bispectrum segments, with the SNR determined from the following equation:
\begin{equation}
\rm{SNR} = \frac{\Sigma^{\it M}_{\it i=1}\rm{amp}_{\it i} cos(\theta_{\it ci} - \theta_{\it c})}{\sqrt{\Sigma^{\it M}_{\it i=1}\rm{amp}^2_{\it i} sin^{2}(\theta_{\it ci}-\theta_{\it c})}},
\end{equation} where amp$_{i}$ is the triple product of the amplitudes, $\theta_{ci}$ is the closure phase for each segment, and $\theta_{c}$ is the averaged closure phase. The analysis was performed within the Haystack Observatory Postprocessing System (HOPS) package, based on the theory developed by \citet{1995AJ....109.1391R}. 

The visibilities were first {\it a priori} calibrated, with a subsequent regularization described below. System temperatures are measured before each VLBI scan and applied to the data. Antenna gains were directly determined from observations of planets at the JCMT and SMT. At the CARMA, relative gains were first estimated using observations in interferometric array mode before each VLBI scan, and the gains were then set to a common flux scale using planet observations at the end of each night \citep{2011ApJ...727L..36F}. A day-to-day systematic trend in CARMA gains due to planet observations,  which led to an apparent brightening of all VLBI sources, was then removed.

The calibrated data were internally reconciled following the ``gain calibration'' method detailed in \citet{2011ApJ...727L..36F}. Briefly, this assumes that (1) the correlated flux density of the CD baseline, which has projected baseline length of $17-33\,k\lambda$ and an equivalent resolution of 8-15\,$\arcsec$, is equal to the CARMA measured total flux density of 10.25 Jy for 1924-292; (2) the low and high band fluxes should be the same statistically; and (3) measured flux densities on SC and SD baselines should be equal, while JC and JD flux densities are not strictly required to be equal due to the lower SNR. The remaining residual errors are believed to be  $\sim$ 5\% and therefore have been added in quadrature to the random errors of the data. 
Figure~\ref{Fig:radplt} shows the variations of the correlated flux density with radial distance from the origin of the uv plane (left) and with time (right).
In Figure~\ref{Fig:clplt}, we show the measured closure phase on the SCD/JCD (left) and SJC/SJD triangles (right). The average of closure phase uncertainties on the SJC/SJD triangle is about $2.6^{\circ}$. The high repeatability of the correlated flux over the three days and reliable measurement of nonzero closure phases indicate a robust detection of a stable source structure. 

\section{Results}
\label{results}

\begin{figure*}[ht]
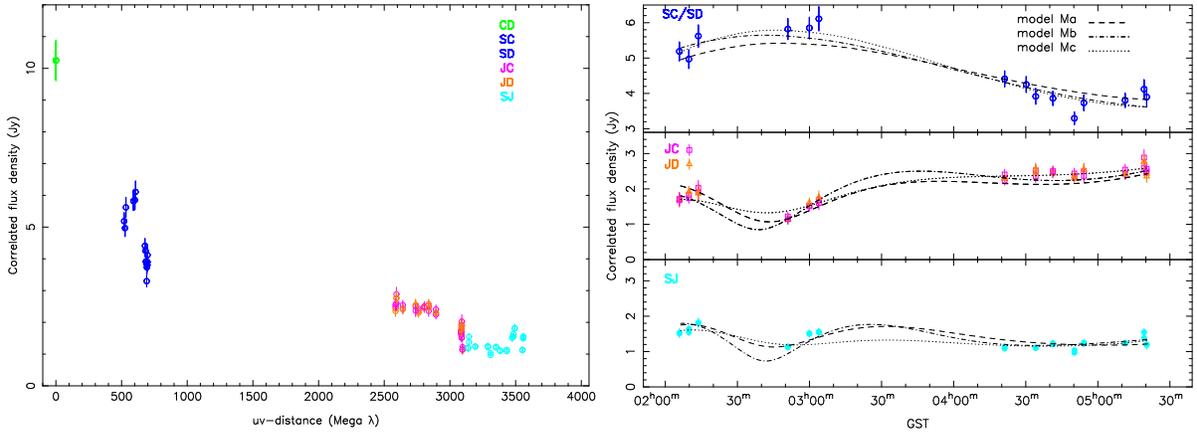

\begin{center}
\includegraphics[angle=-90,width=0.475\textwidth,clip]{fg2a.ps}
\includegraphics[angle=-90,width=0.475\textwidth,clip]{fg2b.ps}
\caption{Left: Correlated flux density as a function of uv distance. Since the source structure is not circularly symmetric, the correlated flux density will change for a given uv distance as the earth rotates. Right: Correlated flux density as a function of time with comparison of the three models.}
\label{Fig:radplt}
\end{center}
\end{figure*}

\begin{figure*}[ht]
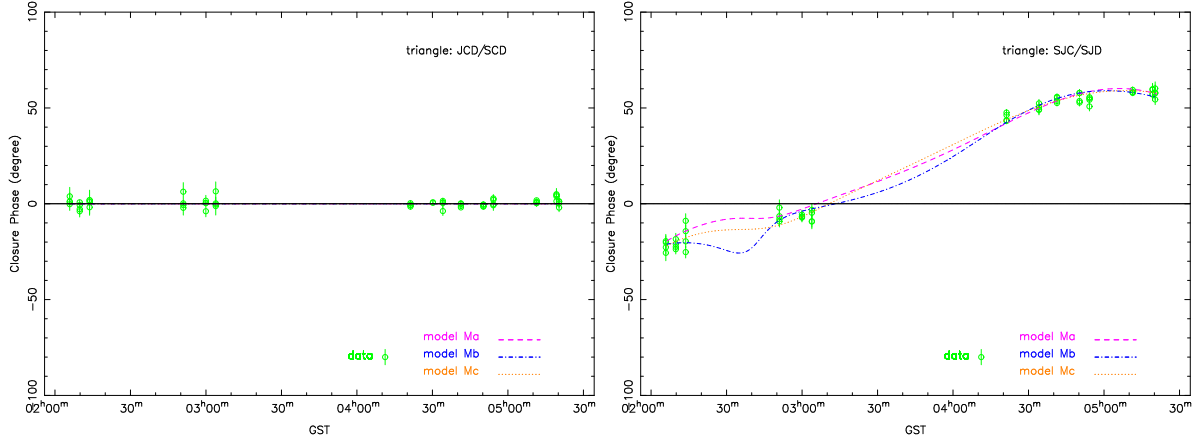

\begin{center}
\includegraphics[width=0.35\textwidth,angle=-90,clip]{fg3a.ps}
\includegraphics[width=0.35\textwidth,angle=-90,clip]{fg3b.ps}
\caption{Plot of closure phase as a function of time for the SCD/JCD triangles (left), and SJC/SJD triangles (right) for the three-day observations. The predicted closure phases for the models are also shown. Note that the expected closure phase for the trivial SCD/JCD triangles is zero.}
\label{Fig:clplt}
\end{center}
\end{figure*}

The calibrated amplitudes and closure phases were modeled with Gaussian functions to parametrize the source structure. 
Least-squares fitting of Gaussian models to VLBI data and error estimation are discussed, e.g., in~\citet{1995ASPC...82..268P}.
The variations evident in the correlated flux densities on the SC and SD baselines (Figure \ref{Fig:radplt}) and the nonzero closure phases (Figure~\ref{Fig:clplt}) indicate that a multicomponent structure is required to fit the data. 

Due to the limited uv coverage, we chose to reproduce the observed flux density by considering two classes of simple models. In the first class (models Ma and Mb), the flux densities on VLBI scales were fitted with two compact Gaussian components, and the drop of correlated flux density of $\sim$ 4 Jy between the CD and SC/SD baselines (Figure \ref{Fig:radplt}) is interpreted as due to an extended component. In this case, a component with a size between the scales probed by the SC/SD (a few hundred microarcseconds) and the CD (a few arcseconds) baselines is required to account for this drop and was fixed during the fitting. However, the position of this component is not constrained by our data because the measured closure phases on SCD/JCD triangles are essentially zero ($0.5^{\circ}\pm2.0^{\circ}$). In model Ma, two circular Gaussians were introduced for the compact flux density, which gives a reduced $\chi^2_{\nu}$=2.0 for all the data (7 parameters, 253 degrees of freedom), indicating that the source structure may be more complicated than assumed. Therefore, the second compact component was replaced by an elliptical Gaussian in model Mb, with reduced $\chi^2_{\nu}=1.5$. In contrast, the second class of model consists of three circular Gaussian components (model Mc), which fits both the VLBI scale and arcsecond scale (CD baseline) flux with three compact components (reduced $\chi^2_{\nu}= 1.2$). With the caveat that the data should not be overinterpreted, we do not attempt to introduce more parameters to obtain a reduced $\chi^2_{\nu}$ even closer to unity. The fitted component flux density, radial distance and position angle with respect to the presumed core, major axis (FWHM) of the Gaussian function, axis ratio, position angle of the major axis, and brightness temperature T$_{b}$ (in the source frame) for the three models are given in Table~\ref{Table:model}. We report uncertainties based on the size of the region around the best-fit point in parameter space corresponding to 68.3\,\% probability. A comparison of model fits to the amplitudes and closure phases for the three-day observations is also shown in Figure~\ref{Fig:radplt} (right) and \ref{Fig:clplt}.

\begin{table*}[htdp]
\caption{Results of Model-Fitting of the compact flux in 1924-292.}
\newsavebox{\tablebox}
\begin{lrbox}{\tablebox}
\begin{threeparttable}[b]
\begin{tabular}{cccccccccc}
\hline
\hline 
Model &$\chi^2_{\nu}$&ID &Flux Density &Distance &P.A.&Major &Ratio &$\phi$&T$_{b}$\\
 && &(Jy) &($\mu$as) &(degree) &($\mu$as)&(-)&(degree)&($10^9$ K)\\
\hline
  Ma&2.0&0&$4.6\pm0.3$& \nodata& \nodata &$>$400-600&1.0&\nodata&$<0.9$\\
  &&1& $2.3\pm0.2$ & 0& 0&$21\pm3$&1.0&\nodata&$160$\\
  &&2&$3.4\pm0.2$ &$371\pm3$&$-55\pm1$&$48\pm2$&1.0&\nodata&$46$\\
\hline
Mb&1.5&0&$3.6\pm0.5$&\nodata&\nodata&$>$400-600&1.0&\nodata&$<0.7$\\
  &&1&$2.5\pm0.2$&0&0&$25\pm3$&1.0&\nodata&$120$\\
  &&2&$4.2\pm0.3$&$363\pm4$&$-56\pm1$&$150\pm19$&$0.3\pm0.1$&$-21\pm2$&$20$\\
\hline
Mc&1.2&0&$3.6\pm0.3$&                        0&                         0&$31\pm2$&1.0&\nodata&120\\
   &&1&$2.0\pm0.2$&$163\pm12$&$-73\pm5$&$50\pm7$&1.0&\nodata&26\\
   &&2&$4.5\pm0.3$&$379\pm7$&$-53\pm1$&$58\pm3$&1.0&\nodata&42\\
\hline
\end{tabular}
\end{threeparttable}
\end{lrbox}
\resizebox{1.0\textwidth}{!}{\usebox{\tablebox}}
\label{Table:model}
\end{table*}%

\section{DISCUSSION}\label{discussion} 
\begin{figure}[ht]
\begin{center}
\includegraphics[width=0.45\textwidth,clip]{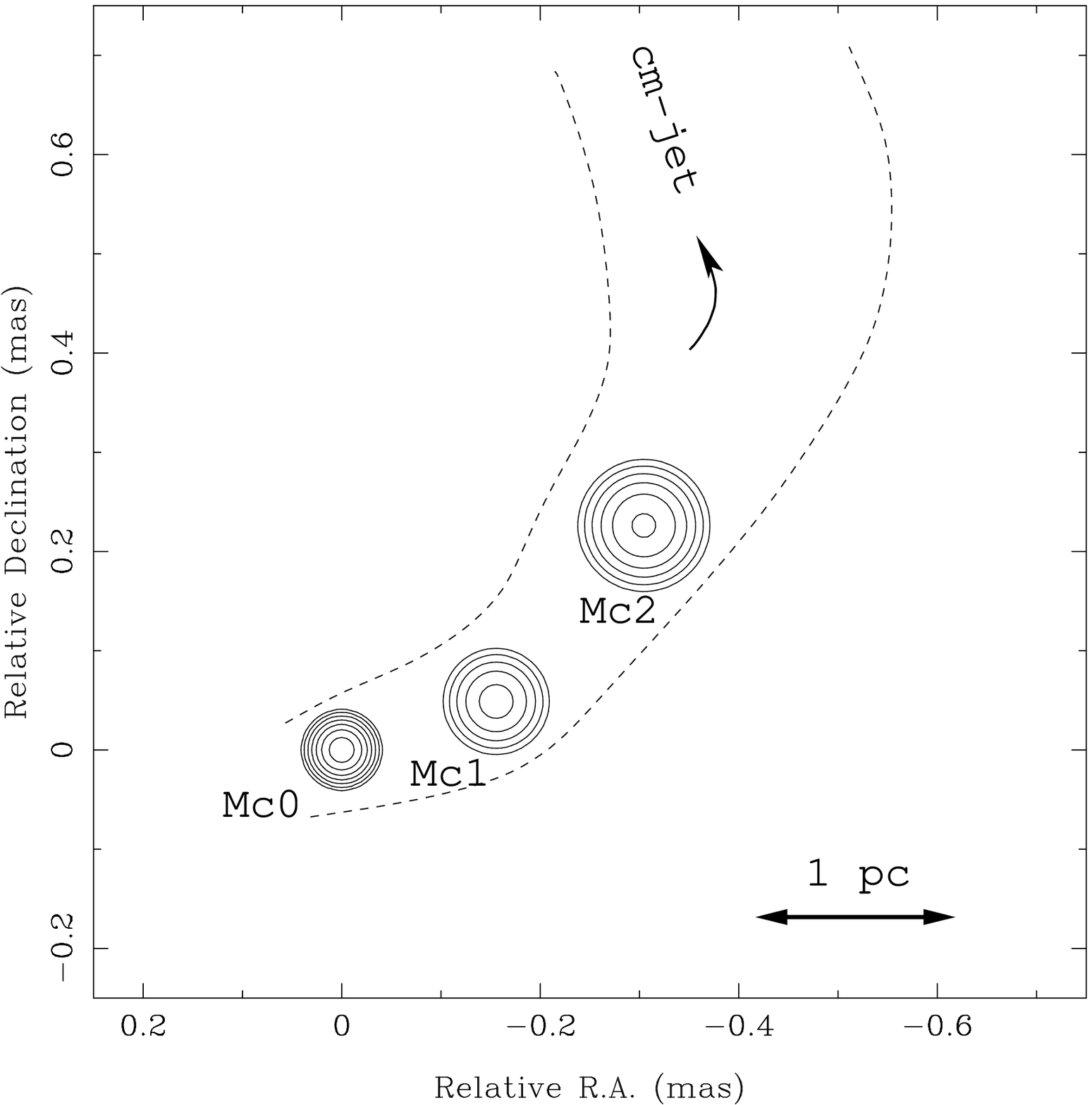}
\caption{Model image of 1924-292. Contours are drawn at 1, 2, 4, ..., 64\,\% of the peak brightness. The two dashed curves indicate (schematically) how the inner jet is bent toward the cm jet.}
\label{Fig:map}
\end{center}
\end{figure}

There is broad consistency among the three models we have considered: They all require two compact components,
$\sim$ 20--30\,$\mu$as (Ma1/Mb1/Mc0) and $\sim$ 50--80\,$\mu$as in size, separated by  $\sim$370\,$\mu$as at a P.A. of $-54^{\circ}$.
These two components are the most reliably measured features by our data, and the accurately measured closure phases remove the $180^{\circ}$ of degeneracy in the jet position angle.

Of the three models, model Mc gives the best fit based on the $\chi^2$-fitting results; however, we cannot completely rule out the first class of models. From Figures~\ref{Fig:radplt} (right) and \ref{Fig:clplt}, it is obvious that some extra data samples on the existing baselines would easily discriminate between the three models even with the present-day sensitivity. Nevertheless, with the most compact component being the core and the sizes of the three components increasing with distance from the core, model Mc is more consistent with the expectation of an expanding jet. Model C also accounts for all the flux in specific components and therefore is preferred. We show its model image in Figure~\ref{Fig:map}. The three-component jet model may be approximating a continuous expanding jet structure, perhaps with nonuniform brightness. Sensitivity and uv coverage limitations currently prevent us from being able to model such a structure, but future observations will be able to directly image the jet structure. 

The fraction of the core flux density $\frac{S_{\rm core}}{S_{\rm total}}$, which indicates the compactness of a source,  is $\sim$ 20\,\% at 1.3\,mm from our observations.  The core, however, is very compact and strong ($\gtrsim$2\,Jy), making it suitable as a fringe finder for 1.3 mm VLBI observations.
It can be seen from Table~\ref{Table:model} and Figure~\ref{Fig:map} that the inner jet structure determined by 1.3 mm VLBI agrees well with the reported inner jet orientation at lower frequencies~\citep[e.g.,][]{2002aprm.conf..401S}. The three-component model at 3.5~mm by \citet{2008AJ....136..159L} does not fit the jet structure seen at 1.3~mm, and the P.A. between their two components away from the core is about $70^{\circ}$. We note, however, that the 3.5~mm observations were made at a different epoch, and the significance of these components is unclear given the limited dynamic range of the 3.5~mm map.

The component (Ma2/Mb2/Mc2) may serve as a link to the extended cm-jet toward a P.A. of $\sim 30^{\circ}$ and can be interpreted as the result of a local bend in the jet toward the observer (Figure~\ref{Fig:map}). Interestingly, the innermost component of about 160\,$\mu$as (0.8 pc) from the assumed core in model Mc (component Mc1) extends to the northwest along a P.A. of $-70^{\circ}$. Its size lies well between the sizes of component Mc0 and Mc2, consistent with the expectation of a fanning-out of the jet. With this component, the innermost jet seems to bend more sharply with respect to the cm emission than previously known with VLBI at lower frequencies, indicating that the jet at these sub-pc scales is extremely curved, reminiscent of a helical jet structure. Furthermore, if this component is associated with the recent mm flare started in 2008\footnote{http://sma1.sma.hawaii.edu/callist/callist.html?plot=1924-292}, and assuming a time lag of 0.1 yr between component ejection and onset of a mm-flare~\citep{2003ASPC..299..249K}, we obtain an estimate of the jet speed of 0.2 mas/yr (4 c), slightly faster than, but still consistent with, the reported jet speed of 3\,c~\citep{2002aprm.conf..401S}. The compact double within the inner 1 pc regions reported by \citet{2000aprs.conf..155S, 2002aprm.conf..401S} may be associated with either Mc0 and Mc1 or Mc0 and Mc2. However, cross identification is difficult because of the long gap in time between these observations. 

The radio core is very compact, and the measured angular size of 31\,$\mu$as translates into a linear size of 0.15\,pc, or $4.6\times10^{17}$\,cm. For this component, we obtained a source-frame brightness temperature of $1.2\times10^{11}$K, less than the inverse Compton limit \citep{1969ApJ...155L..71K} and the equipartition limit \citep{1994ApJ...426...51R}. Therefore, the measured core brightness temperature alone does not readily imply relativistic beaming, although the reported superluminal motion is suggestive of a beaming effect. Indeed, a very large Doppler factor (> 80) was reported in this source \citep{1999PASJ...51..537F}. 

The core brightness temperature at 1.3\,mm is well below those measured at cm wavelengths \citep[$\gtrsim 3 \times 10^{12}$ K,][and references therein]{1999PASJ...51..513S}. This is even true when we consider those derived brightness temperatures as lower limits because the core is unresolved at cm wavelengths. On the one hand, this may suggest that relativistic beaming does not play a significant role in the core region at 1.3\,mm. One possibility is that the jet bends such that the Doppler beaming effect is not maximized at the position of the core at 1.3\,mm, but somewhere downstream of the jet.  Another possibility is that the jet undergoes parsec-scale acceleration, so that the jet is gradually beamed at some distance from the central engine, and the inner jet traced at 1.3\,mm is still accelerating. On the other hand, our observations are at frequencies above the turnover frequency of the synchrotron jet emission for this source. The self-similarity of the parsec-scale jet is believed to break down in the regions probed at these high frequencies~\citep{2009arXiv0909.2576M}. 
Compared with previous low-frequency observations, we probed ``deeper'' into the core region, where the brightness temperature can be intrinsically lower. This supports the decelerating jet model or particle-cascade models, as discussed by \citet{1995PNAS...9211439M}, which predict a lower brightness temperature for the inner jet close to its origin.

\section{Summary and Future Prospects.}
We have presented the first high-resolution VLBI observations of the quasar 1924-292 at 230\,GHz. For the first time, an AGN jet is spatially resolved with 1.3\,mm VLBI with robust closure phase measurements. At pc scales, the inner jet direction is oriented toward the northwest (P.A.= $-53^{\circ}$) with respect to the assumed core, consistent with previous results. The innermost jet appears to bend more sharply than previously known, and the jet curvature seen at 1.3\,mm seems to be more pronounced than at lower frequencies.

The compact core has a linear size of 0.15\,pc and a brightness temperature of $1.2\times10^{11}$K,  much lower than previous measured values at lower frequencies. This may indicate that the self-similar jet breaks down at these probed scales and at frequencies at which the jet becomes optically thin. VLBI observations at 1.3\,mm are beginning to discriminate between inner jet models, and a decrease of brightness temperature at higher frequencies (and therefore closer to the jet origin) provides evidence in support of the decelerating jet model or particle-cascade models.

 Further improvement of the array performance, e.g., by adding suitable VLBI sites and increasing bandwidth recording rate and station collecting area via phased array techniques, will allow the EHT to directly image the vicinity of super massive black holes at the Galactic center and nearby galactic nuclei with a spatial resolution of a few Schwarzschild radii in the foreseeable future. The robust measurement of closure phase for 1924-292 opens the way for detecting and constraining flare structures on Event-horizon scales for the Galactic center~\citep{2009ApJ...695...59D}.

\acknowledgments
We thank the referee for useful comments and suggestions.
High-frequency VLBI work at MIT Haystack Observatory is supported by grants from the National Science Foundation (NSF). 
The Arizona Radio Observatory (ARO) is partially supported through the NSF University Radio Observatories (URO) program under grant no. AST 1140030. The Submillimeter Array is a joint project between the Smithsonian Astrophysical Observatory and
the Academia Sinica Institute of Astronomy and Astrophysics and is funded by the Smithsonian Institution and the Academia Sinica. 
Funding for ongoing CARMA development and operations is supported by the NSF and the CARMA partner universities. 

 {\it Facilities:} \facility{CARMA}, \facility{ARO SMT},
 \facility{JCMT}, \facility{SMA}

 \end{document}